\documentclass[10pt,conference]{IEEEtran}
\IEEEoverridecommandlockouts
\usepackage{cite}
\usepackage{amsmath,amssymb,amsfonts}
\usepackage{graphicx, physics, soul, hyperref}
\usepackage{textcomp}
\usepackage{xcolor, algorithm, algpseudocode}
\def\BibTeX{{\rm B\kern-.05em{\sc i\kern-.025em b}\kern-.08em
    T\kern-.1667em\lower.7ex\hbox{E}\kern-.125emX}}
\begin{document}

\title{Non-parametric Greedy Optimization of Parametric Quantum Circuits
}

\author{\IEEEauthorblockN{Koustubh Phalak}
\IEEEauthorblockA{\textit{Computer Science \& Engineering Dept.} \\
\textit{Pennsylvania State University}\\
State College, PA. \\
krp5448@psu.edu}
\and
\IEEEauthorblockN{Swaroop Ghosh}
\IEEEauthorblockA{\textit{School of EECS} \\
\textit{Pennsylvania State University}\\
State College, PA \\
szg212@psu.edu}
}

\maketitle

\begin{abstract}
The use of Quantum Neural Networks (QNN) that are analogous to classical neural networks, has greatly increased in the past decade owing to the growing interest in the field of Quantum Machine Learning (QML). A QNN consists of three major components: (i) data loading/encoding circuit, (ii) Parametric Quantum Circuit (PQC), and (iii) measurement operations. Under ideal circumstances the PQC of the QNN trains well, however that may not be the case for training under quantum hardware due to presence of different kinds of noise. Deeper QNNs with high depths tend to degrade more in terms of performance compared to shallower networks. This work aims to reduce depth and gate count of PQCs by replacing parametric gates with their approximate fixed non-parametric representations. We propose a greedy algorithm to achieve this such that the algorithm minimizes a distance metric based on unitary transformation matrix of original parametric gate and new set of non-parametric gates. From this greedy optimization followed by a few epochs of re-training, we observe roughly 14$\%$ reduction in depth and 48$\%$ reduction in gate count at the cost of 3.33$\%$ reduction in inferencing accuracy. Similar results are observed for a different dataset as well with different PQC structure. 
\end{abstract}


\section{Introduction}
Quantum Computing (QC) is a rapidly growing field that constantly keeps on evolving due as a results of new and significant findings from researchers in the domain. In just a decade, the scope of quantum computing has shifted from simulation \cite{georgescu2014quantum,buluta2009quantum} all the way to quantum utility \cite{kim2023evidence} that ensures reliable quantum operations even under noise. This has opened up path for many subfields in quantum computing along with exsisting fields such as Quantum Machine Learning (QML) \cite{schuld2015introduction}, quantum chemistry \cite{cao2019quantum}, finance \cite{herman2022survey}, healthcare \cite{ur2023quantum}, and security \cite{saki2021survey}. For QML in particular, the growth has been very fast due to concurrent advances in both classical machine learning as well as quantum computing \cite{zhang2020recent}. Quantum Neural Networks (QNN), which are quantum analogous models of classical neural networks have emerged as a promising class of QML models. Typically, the QNN has three segments: (i) the data loading stage for loading classical data, (ii) Parametric Quantum Circuit (PQC) that consists of trainable rotation gates along with entangling gates to train the QNN, and (iii) measurement operations to measure quantum states classically for optimization. Typically, QNNs use traditional classical optimization techniques such as Adam, Stochastic Gradient Descent (SGD), Alternating Direction Method of Multipliers (ADMM), adaptive learning rate, etc. \cite{sun2019survey} to train and update the rotation angles of the parametric gates.

PQCs generally contains a layered structure where multiple layers of fixed set of rotation and entangling gates (also referred to as ansatz) are used. This is analogous to classical neural networks having multiple layers of neurons. As an example, Pennylane library from Xanadu has basic entangler layer and strongly entangling layer from \cite{schuld2020circuit}, simplified 2-design layer from \cite{cerezo2021cost}, and continuous-variable (CV) QNN layers from \cite{killoran2019continuous} pre-defined, which one can add repetitively in their custom quantum circuit. The addition of such layers in the PQC is said to make the overall QNN to be having higher expressive power compared to classical neural networks \cite{abbas2021power, du2020expressive}. This implies that QNNs have a strong potential to demonstrate quantum advantage.

Under ideal, noiseless conditions a QNN trains well. Realistically however, there are two cruicial challenges: 
\textbf{(i) Susceptibility to noise:} modern Noisy Intermediate Scale Quantum (NISQ) devices are plagued with plethora of noise such as decoherence (unwanted interaction of the quantum system with external environment), gate errors (erroneous gate operation implementations), readout errors (inaccurate measurement apparatus) and crosstalk errors (unwanted interation of neighboring/coupled qubits). These erroneous sources of noise degrade performance of QNNs. Furthermore, deeper QNNs with more layers promises to learn better but often face higher degradation compared to shallower networks due to larger depth and gate count. \textbf{(ii) Quantum hardware constraints:} Generally, a quantum computer does not contain all quantum gates native to it, but rather a set of basis gates that is typically a universal set. According to the Solovay-Kitaev theorem \cite{dawson2005solovay}, a universal set allows \textit{decomposition} of any non-native gate as an approximate sequence of native gates present in the basis gate set. Additionally, not every qubit is connected to every other qubit in the quantum hardware. There is limited connectivity between qubits which is respresented by coupling map of the quantum hardware. Due to such \textit{coupling constraints}, any multi-qubit operations between physically disconnected qubits require SWAP gate operations to bring the quantum states between physically connected qubits. The combined process of decomposition and SWAP insertion is often referred to \textit{transpilation} and is an important step for logical-to-physical qubit mapping. The above hardware constraints i.e., limited native gate set and qubit connectivity result in increased gate count and circuit depth when mapped to real hardware degrading the fidelity of the computation.

In this work, we focus only on the decomposition problem and propose a greedy algorithm-based optimization of PQCs that is able to reduce post-decomposition depth and gate count. We specifically optimize only parametric rotation gates in the PQC by replacing them with fixed non-parametric gates that potentially have lesser decomposition depth compared to the original parametric gate. We greedily optimize a unitary transformation matrix-based distance metric that minimizes the distance between the transformation matrices of the original gate and selected set of non-parametric gates. The original parametric gate is then replaced with this set of non-parametric gates in the original PQC, and this action is repetitively performed for all the parametric gates. The optimized QNN may reduce the inference performance compared to the original QNN, so we re-train the new QNN for a few epochs to regain the original performance. All the code corresponding this work can be found in our GitHub repository\footnote{GitHub repository link: \href{https://github.com/KoustubhPhalak/Greedy-PQC-Optimization}{Greedy-PQC-Optimization}}. 

In the rest of the paper, Section II introduces general background on quantum computing along with related works on QNN compression methods and quantum circuit transformation techniques. Section III presents the proposed greedy algorithm to perform non-parametric approximation of parametric rotation gates, and Section IV applies this algorithm on pre-trained PQCs to reduce overall post-decomposition depth and gate count. We also analyze inferencing on optimized PQC after re-training. Finally, Section V provides general discussion on the proposed greedy algorithm and Section VI concludes the paper.

\section{Background and Related Works}
\subsection{Quantum Computing}
Analogous to bits used in classical computing, quantum computing utilizes quantum bits (qubits in short) to store information in quantum Hilbert space. This information is often represented in the form of quantum state.
It is mathematically denoted as
$\ket{\psi}=$
$\big[\begin{smallmatrix}
\alpha \\
\beta
\end{smallmatrix}\big]$, where $|\alpha|^2$ is the probability of qubit being measured to 0 and $|\beta|^2$ is the probability of qubit being measured to 1. There are two special states 
$\ket{0}=$
$\big[\begin{smallmatrix}
1 \\
0
\end{smallmatrix}\big]$ and
$\ket{1}=$
$\big[\begin{smallmatrix}
0 \\
1
\end{smallmatrix}\big]$ which are known as basis states. The quantum state of a qubit can be changed with the help of quantum mechanical phenomena such as superposition, entanglement and interference. Quantum gates, which are unitary matrix operations are used to bring about this change of quantum state. These gates either work on single or multiple qubits. An ordered set of such quantum gate operations is a quantum circuit. All the gate operations of quantum circuit are followed by measurement operation that causes wavefunction collapse \cite{von2018mathematical} and measures the quantum state into a binary classical value. A special kind of quantum circuit is Parametric Quantum Circuit (PQC), which consists of parametric rotation gates. Typically, PQCs are trainable circuits which can be trained like an ML model using classical optimization techniques, and are an integral part of QNNs.

\subsection{Noise in Quantum Hardware}
Various kinds of noise can adversely affect the performance of QNNs. Decoherence error occurs when a qubit or multiple qubits interact with external environment which leads to loss of energy from the qubits. Crosstalk error is similar to decoherence with the difference being that unwanted interaction can happen between qubits that are coupled to each other. Quantum gates are typically implemented using microwave (superconducting qubits) or laser (trapped-ion qubits) pulses. There can be errors in application of these pulses which can lead to gate errors. Finally, there are different implementations of measurement operation such as photon detectors (photonic qubits), flourescence intensity (trapped-ion qubits) or resonator coupling (superconducting qubits). Once again, imprecise measurements or inaccuracies in the apparatus can lead to readout errors.

\subsection{QNN Compression Techniques}
QNN compression approaches imitate classical neural network compression techniques for quantum circuits. \cite{alam2023knowledge} proposes knowledge distillation \cite{hinton2015distilling} for PQCs which uses a unitary transformation matrix-based approximate synthesis of PQCs to reduce the overall depth and gate count of the quantum circuit. The distinctions with our methodology is as follows. Firstly, the approximate synthesis \cite{alam2023knowledge} still yields parametric gates unlike our work which yields only non-parametric gates and second, the authors use simulated annealing to perform optimization whereas we use greedy algorithm to obtain gate optimization. In another related work \cite{hu2022quantum}, authors propose quantization and pruning techniques for QNNs. Based on the native basis gate set of the quantum hardware, the authors create lookup table of decomposition length for different angles for each single-qubit parametric gate. Based on the angle, if (i) the decomposition depth leads to zero i.e. an identity gate (\textit{+I} or \textit{-I}), then the gate can be pruned. (ii) depending on the quantized rotation angle, the decomposition depth of the parametric gate can be reduced. For example, in a basis set of $\{CX,ID,RZ,SX,X\}$, an RX gate with angle $\theta$ can be decomposed as 

\small\[RX(\theta)=RZ \left( \frac{5\pi}{2} \right)\cdot SX\cdot RZ(\theta+\pi)\cdot SX\cdot RZ\left(\frac{\pi}{2}\right)\]

For $\theta=0$ $\xrightarrow{}$ $RX(0)=I$ (pruned), $\theta=\frac{3\pi}{2}$ $\xrightarrow{}$ $RX\left(\frac{3\pi}{2} \right)=RZ(-\pi)\cdot SX \cdot RZ(-\pi)$ (3 depth), $\theta=\frac{\pi}{2}$ $\xrightarrow{}$ $RX\left(\frac{3\pi}{2} \right)=SX$ (1 depth). Thus, quantizing to certain angles leads to compression of gates in the QNN. The authors employ alternating direction method of multipliers (ADMM) optimization method on the lookup table to come up with the optimal decomposition on each quantized parametric gate. However, this problem suffers from an exponential search space and requires a reduction of the lookup table to reduce the optimization time.

\begin{algorithm}[t]
\caption{Parametric Gate Transformation Algorithm}\label{alg:param_gate_greedy}
\begin{algorithmic}[1]
\Function{ParamGateTransform}{$qc\_original$, $N$, $k$}
    \State Initialize $qc\_new$ with 1 qubit
    \State Compute $tm\_original$ of $qc\_original$
    \State $single\_qubit\_gates = [x, y, z, h, s, t, id,$
    \Statex\hspace{\algorithmicindent}$\sqrt{x}, s^{\dag}, \sqrt{x}^{\dag}, t^{\dag}]$
    \State $final\_dist=$ 1000
    \State $prev=$ `null' (stands for previously selected gate)
    \For{$i$ from 1 to $N$}
        \State $dist\_dict = \{\}$ 
        \For{each gate in $single\_qubit\_gates$}
            \State $dist\_dict[gate]=$1000
        \EndFor
        \For{each gate in $single\_qubit\_gates$}
            \If{$gate==prev$}
                \State Continue
            \EndIf
            \State $qc\_new$.add($gate$)
            \State Compute $tm\_new$ of $qc\_new$
            \State $dist\_dict[gate]=1-\frac{Tr(tm\_new^{\dag}\cdot tm\_original)}{dim(tm\_new)}$
            \State $qc\_new$.delete($gate$)
        \EndFor
        \State sort($dist\_dict$,by value)
        \State $top\_k\_gates=dist\_dict.keys()[:k]$ 
        \State $best\_gate=random\_choice(top\_k\_gates)$
        \If{$final\_dist>dist\_dict[best\_gate]$}
            \State $prev=best\_gate$
            \State $final\_dist=dist\_dict[best\_gate]$
            \State $qc\_new$.add($best\_gate$)
        \Else
            \State Continue
        \EndIf
    \EndFor
    \State \Return $final\_dist$, $qc\_new$
\EndFunction
\end{algorithmic}
\end{algorithm}

\subsection{Quantum Circuit Transformation Techniques}
Another similar problem that is being solved is quantum circuit transformation, where the goal is to find a functionally equivalent circuit of the original circuit for better performance on hardware. There are two main approaches for quantum circuit transformation: (i) \textit{rule-based transformation} that uses a set of gate cancellation rules on sub-circuits to perform circuit transformation. These are typically incorporated in transpilers such as Qiskit \cite{aleksandrowicz2019qiskit}, Quilc 
\begin{algorithm}[t]
\caption{Optimization of Parametric Quantum Circuit (PQC) using ParamGateTransform() function}\label{alg:pqc_optimization}
\begin{algorithmic}[1]
\Function{OptimizePQC}{PQC, tolerance}
    \State Initialize an empty quantum circuit $qc\_new$ with the same number of qubits as PQC
    \State Parse the PQC and create a list $gates\_list$ with gate metadata
    \For{each gate $g$ in $gates\_list$}
        \If{$g$ is an RX/RY/RZ gate}
            \State Create new single qubit circuit $qc\_temp$ and 
            \Statex\hspace{\algorithmicindent}\hspace{\algorithmicindent}\hspace{\algorithmicindent}add $g$
            \State $dist$, $qc\_opt$ $=$ \Statex\hspace{\algorithmicindent}\hspace{\algorithmicindent}\hspace{\algorithmicindent}ParamGateTransform($qc\_temp$, $20$, $4$)
            \If{$dist$ $<$ tolerance}
                \State $qc\_new$.add($qc\_opt$)
            \Else
                \State $qc\_new$.add($qc\_temp$)
            \EndIf
        \ElsIf{$g$ is a CNOT gate}
            \State $qc\_new$.add(CNOT)
        \EndIf
    \EndFor
    \State \Return $qc\_new$
\EndFunction
\end{algorithmic}
\end{algorithm}
\cite{mark_skilbeck_2020_3967926} and $\ket{tket}$ \cite{sivarajah2020t}; (ii) \textit{search-based transformation} that is more flexible and tries to find a functionally equivalent quantum circuit within a search space. Typically, reinforcement learning is used to aid for this method \cite{li2023quarl,fosel2021quantum, moflic2023graph} while some other methods also include usage of Monte-Carlo tree search \cite{zhou2020monte} and ZX-Calculus \cite{duncan2020graph}. Recently, QuantumNAS \cite{wang2022quantumnas} was proposed, that trains a SuperCircuit which is then used to optimize SubCircuits with respect to target qubit mapping of a quantum hardware.
\section{Proposed Greedy Algorithm}
We propose a greedy optimization algorithm to individually optimize a parametric gate to an approximation of sequence of non-parametric gates. For each step of the greedy algorithm, we minimize a distance metric that depends on the unitary transformation matrices of the original gate and the new approximation of non-parametric gates. 
If $U$ is the unitary transformation matrix representation of the original parametric gate and $V$ is the unitary transformation matrix representation of the set of non-parametric gates, then the distance metric \cite{alam2023knowledge} is defined as $d = 1 - \frac{Tr(V^{\dagger}U)}{dim(V)}$, where $Tr()$ is the trace function and $dim()$ is the number of dimensions of a matrix. In an ideal scenario, $V^{\dag}U=I_N$ where $I_N$ is $N$-dimensional identity matrix. So, $Tr(I_N) = N = dim(V)$, leading to $d = 1 - 1 = 0$. In other words, a lower distance will reduce the gap between transformation matrix representations $U$ and $V$. 
Note that the size of the transformation matrix depends exponentially on the number of qubits used ($2^n * 2^n$ for $n$ qubits). However, as we shall show, we bypass this exponential requirement by performing the greedy algorithm only on individual single-qubit parametric gates, which reduces the size of transformation matrices only to size $2*2$.

\begin{figure*}[t]
    \centering
    \includegraphics[width=0.85\linewidth]{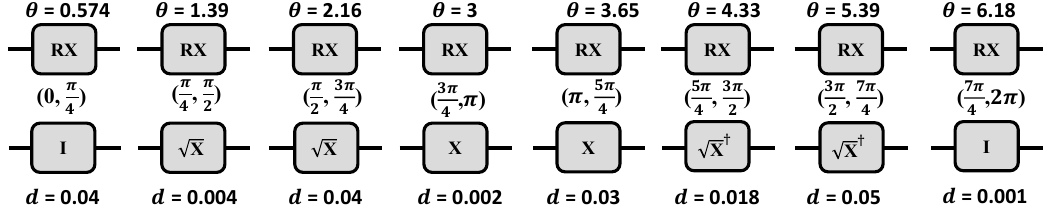}
    \caption{RX gate approximations for different angle values}
    \label{fig:rx_example}
\end{figure*}

The proposed greedy approach is presented in Algorithm \ref{alg:param_gate_greedy}. We represent the original gate as a single qubit circuit $qc\_original$ with transformation matrix $tm\_original$, and new approximation in another single qubit circuit $qc\_new$ with overall transformation matrix $tm\_new$. We select the following list of single qubit non-parametric gates (shown in $single\_qubit\_gates$ list) to create our approximations: $x$, $y$, $z$, $h$, $s$, $t$, $id$, $\sqrt{x}$, $s^{\dag}$, $\sqrt{x}^{\dag}$, and $t^{\dag}$. For $N$ iterations, we use brute-force method for all the specified single-qubit non-parametric gates by (i) adding the gate to $qc\_new$, (ii) computing subsequent $tm\_new$, (iii) computing the distance between $tm\_original$ and $tm\_new$, and finally (iv) removing the non-parametric gate from $qc\_new$. The distance associated with each parametric gate is stored in a dictionary $dist\_dict$, which at the end is sorted based on distance values in ascending order, and one gate is randomly selected among top $k$ choices. This selected gate is the $best\_gate$. Finally, an overall distance check is done between the overall circuit distance $final\_dist$ and the $best\_gate$ distance $dist\_dict[best\_gate]$. If the distance provided by the addition of $best\_gate$ is lower compared to $final\_dist$ only then will $best\_gate$ be added to $qc\_new$, otherwise it will not be added. Furthermore, $prev$, which stores the previously added gate will be updated to $best\_gate$ in the scenario that gate is added.

Note that (i) we do not use $prev$ gate in the subsequent iteration for distance calculation to avoid having same gates that cancel out to give identity operation ($x.x=y.y=z.z=h.h=id$) and (ii) we use random choice among $k$ least distance gates in order to avoid scenarios where alternately always the same gate is selected that always cancel out to give identity operation ($s.s^{\dag}.s.s^{\dag}...=t.t^{\dag}.t.t^{\dag}...=id$). For our work, we find out by trial and error that $N=20$ and $k=4$ are the optimal parameter choices for our greedy algorithm. As an example, we show in Fig. \ref{fig:rx_example} how the RX gate is approximated for different angle values ranging from $0$ to $2\pi$ in intervals of $\frac{\pi}{4}$.

\begin{figure}[t]
    \centering
    \includegraphics[width=\linewidth]{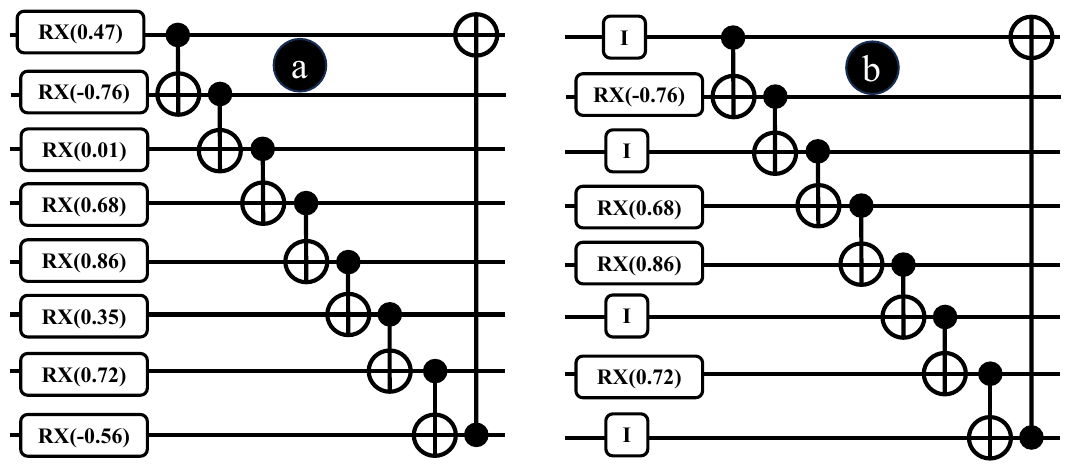}
    \caption{(a) First basic entangler layer of PQC trained on iris dataset, (b) same layer after performing optimization shown in Algorithm \ref{alg:pqc_optimization}}
    \label{fig:iris_first_layer_opt}
\end{figure}

\section{Application on Trained PQCs}
We apply the proposed greedy algorithm on two trained PQCs. The first is an 8-qubit PQC that is trained on classifying UCI Iris dataset \cite{misc_iris_53} and second is a 10-qubit PQC trained on classifying UCI Digits dataset \cite{misc_optical_recognition_of_handwritten_digits_80} (only 0 and 1 digits). For both the PQCs, we use five basic entangler layers \cite{schuld2020circuit} (example is shown in Fig. \ref{fig:iris_first_layer_opt}(a)), that consists of a rotation gate on each qubit (in this case RX gate) followed by CNOT gates connected in a circular fashion. For each dataset, we use Adam optimizer, learning rate of $10^{-3}$, and 50 epochs for training. Later, for additional analysis we also perform optimization on PQCs containing strongly entangling layers as well.

\textbf{Basic entangler layer-based PQC:} A summary of the PQC optimization algorithm is shown in Algorithm \ref{alg:pqc_optimization}. In order to optimize a PQC, we perform following steps: (i) we create an empty circuit with same number of qubits as the original PQC, (ii) we parse the PQC and create a list of gates containing important metadata (qubits operating on, rotation angle, etc.), (iii) we go through the parsed list of gates one by one and then if we encounter an RX gate we perform greedy optimization. We set a tolerance value for the distance metric, and if the distance obtained is lower than the tolerance value the original RX gate is replaced with the non-parametric approximation and added to the empty circuit, else the original RX gate will be added, and (iv) if we encounter a CNOT gate it is added as is without making any changes.

\begin{table}[t]
\centering
\caption{Tolerance sweep results.}
\label{tab:tolerance_sweep}
\begin{tabular}{|llll|}
\hline
\multicolumn{4}{|c|}{\textbf{Basic Entangler Layer-based PQC}}                                                                                                                                                                                        \\ \hline
\multicolumn{1}{|l|}{\textbf{\begin{tabular}[c]{@{}l@{}}Tolerance\\ value\end{tabular}}} & \multicolumn{1}{l|}{\textbf{Depth}}                     & \multicolumn{1}{l|}{\textbf{Gate count}}                 & \textbf{Accuracy}                     \\ \hline
\multicolumn{4}{|c|}{Iris (66 depth, 240 gate count, 90\% acc)}                                                                                                                                                                              \\ \hline
\multicolumn{1}{|l|}{{\color[HTML]{000000} \textit{\textbf{0.1}}}}                       & \multicolumn{1}{l|}{{\color[HTML]{000000} \textbf{48}}} & \multicolumn{1}{l|}{{\color[HTML]{000000} \textbf{77}}}  & {\color[HTML]{000000} \textbf{33.33}} \\ \hline
\multicolumn{1}{|l|}{{\color[HTML]{000000} \textit{0.01}}}                               & \multicolumn{1}{l|}{{\color[HTML]{000000} \textit{59}}} & \multicolumn{1}{l|}{{\color[HTML]{000000} \textit{206}}} & {\color[HTML]{000000} \textit{86.67}} \\ \hline
\multicolumn{1}{|l|}{\textit{0.001}}                                                     & \multicolumn{1}{l|}{65}                                 & \multicolumn{1}{l|}{224}                                 & 90                                    \\ \hline
\multicolumn{4}{|c|}{Digits (76 depth, 300 gate count, 90.2\% acc)}                                                                                                                                                                          \\ \hline
\multicolumn{1}{|l|}{{\color[HTML]{000000} \textit{\textbf{0.1}}}}                       & \multicolumn{1}{l|}{{\color[HTML]{000000} \textbf{58}}} & \multicolumn{1}{l|}{{\color[HTML]{000000} \textbf{116}}} & {\color[HTML]{000000} \textbf{65.27}} \\ \hline
\multicolumn{1}{|l|}{{\color[HTML]{000000} \textit{0.01}}}                               & \multicolumn{1}{l|}{{\color[HTML]{000000} \textit{64}}} & \multicolumn{1}{l|}{{\color[HTML]{000000} \textit{193}}} & {\color[HTML]{000000} \textit{88.88}} \\ \hline
\multicolumn{1}{|l|}{\textit{0.001}}                                                     & \multicolumn{1}{l|}{73}                                 & \multicolumn{1}{l|}{228}                                 & 90.2                                  \\ \hline
\end{tabular}
\end{table}

An example of the PQC optimization algorithm is shown in Fig. \ref{fig:iris_first_layer_opt}, which shows first basic entangler layer of PQC that was trained for classiying iris dataset. Fig. \ref{fig:iris_first_layer_opt}(a) shows the layer prior to performing the greedy optimization and Fig. \ref{fig:iris_first_layer_opt}(b) shows the same layer after performing the optimization. We can observe that some RX gates that have rotation angle in the range $(0,\frac{\pi}{4})$ have been replaced with identity gate. This observation is in line with our expectations as we can see from Fig. \ref{fig:rx_example}. For this particular example, we have set the distance tolerance value to 0.05. In general, we sweep the tolerance value (to 0.1, 0.01, and 0.001) and perform inferencing on each optimized circuit to obtain testing accuracy. The results of this sweep are tabulated in Table \ref{tab:tolerance_sweep}. 
From the results, we observe that (i) lower tolerance values are more accurate but perform less optimization, (ii) higher tolerance values perform high optimization but suffer from severe performance degradation. We particularly format 0.1 value parameters in bold to denote that it gives low inferencing accuracy but does not have any RX rotation gates left to further re-train the circuit for improvement. We also format 0.01 value parameters in italics to show good amount of circuit optimization for minimal performance degradation, which can be improved since it has RX gates left for training.

\begin{table}[t]
\centering
\caption{Parameter values for optimal tolerance value.}
\label{tab:optimal_values}
\begin{tabular}{|llll|}
\hline
\multicolumn{4}{|c|}{\textbf{Basic Entangler Layer-based PQC}}                                                                                                                                                                                                                                                                                                   \\ \hline
\multicolumn{1}{|l|}{}                                                                 & \multicolumn{1}{l|}{\textbf{\begin{tabular}[c]{@{}l@{}}Original\\ circuit\end{tabular}}} & \multicolumn{1}{l|}{\textbf{\begin{tabular}[c]{@{}l@{}}Optimized\\ circuit\end{tabular}}} & \textbf{\begin{tabular}[c]{@{}l@{}}Optimized \\ circuit re-trained\end{tabular}} \\ \hline
\multicolumn{4}{|c|}{Iris dataset (tolerance = 0.05)}                                                                                                                                                                                                                                                                                                   \\ \hline
\multicolumn{1}{|l|}{\textit{\begin{tabular}[c]{@{}l@{}}Accuracy\\ (\%)\end{tabular}}} & \multicolumn{1}{l|}{90}                                                                  & \multicolumn{1}{l|}{50}                                                                  & {\color[HTML]{000000} \textbf{86.67 (3.33\% red.)}}                              \\ \hline
\multicolumn{1}{|l|}{\textit{Depth}}                                                   & \multicolumn{1}{l|}{66}                                                                  & \multicolumn{1}{l|}{57}                                                                   & {\color[HTML]{000000} \textit{57 (13.6\% red.)}}                                 \\ \hline
\multicolumn{1}{|l|}{\textit{Gate count}}                                              & \multicolumn{1}{l|}{240}                                                                 & \multicolumn{1}{l|}{125}                                                                  & {\color[HTML]{000000} \textit{125 (47.9\% red.)}}                                \\ \hline
\multicolumn{4}{|c|}{Digits dataset (tolerance = 0.06)}                                                                                                                                                                                                                                                                                                 \\ \hline
\multicolumn{1}{|l|}{\textit{\begin{tabular}[c]{@{}l@{}}Accuracy\\ (\%)\end{tabular}}} & \multicolumn{1}{l|}{90.2}                                                                & \multicolumn{1}{l|}{69.4}                                                                 & {\color[HTML]{000000} \textit{95.3 (5.1\% inc.)}}                                \\ \hline
\multicolumn{1}{|l|}{\textit{Depth}}                                                   & \multicolumn{1}{l|}{76}                                                                  & \multicolumn{1}{l|}{60}                                                                   & {\color[HTML]{000000} \textit{60 (26.6\% red.)}}                                 \\ \hline
\multicolumn{1}{|l|}{\textit{Gate count}}                                              & \multicolumn{1}{l|}{300}                                                                 & \multicolumn{1}{l|}{121}                                                                  & {\color[HTML]{000000} \textit{121 (59.6\% red.)}}                                \\ \hline
\end{tabular}
\end{table}

This implies that the optimal tolerance value lies between 0.01 and 0.1. From trial and error, we find out that for bare minimum number of RX gates needed for good re-training performance iris PQC can accomodate a maximum tolerance value of \textit{0.05} while digits PQC can accomodate a maximum tolerance value of \textit{0.06}. Once these tolerance values are selected, we re-train these PQCs for roughly 30-40$\%$ of original number of epochs i.e. 15-20 extra epochs. The re-trained parameter values have been tabulated in Table \ref{tab:optimal_values}. From this table we note that the optimized circuit without re-training leads to significant performance degradation, which is minimized after re-training. For iris dataset, we don't quite reach the original testing accuracy. However for digits dataset we gain a 5.1$\%$ boost in testing accuracy. 

\begin{figure}[t]
    \centering
    \includegraphics[width=\linewidth]{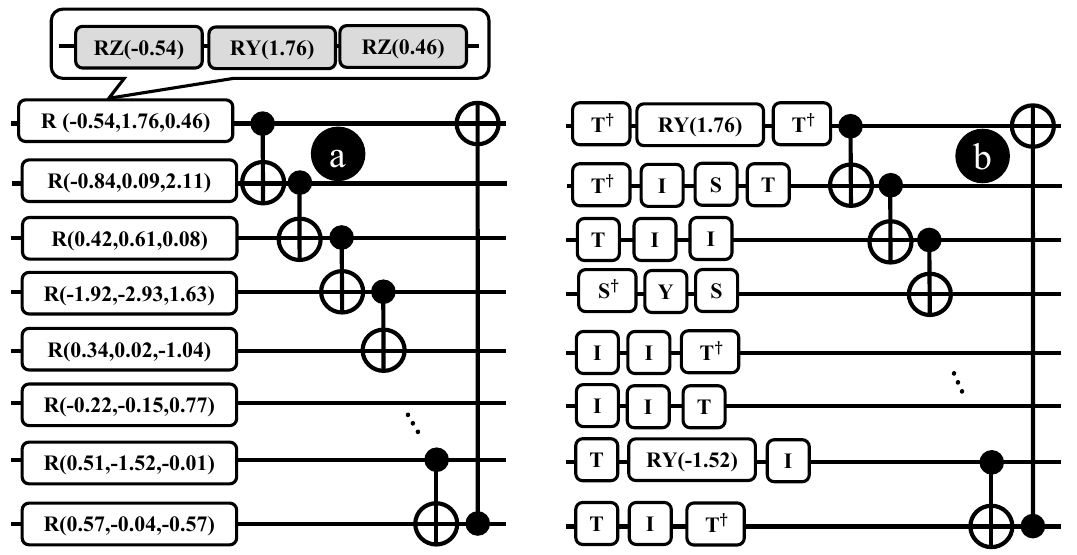}
    \caption{(a) First strongly entangling layer of PQC trained on iris dataset, (b) same layer after performing optimization shown in Algorithm \ref{alg:pqc_optimization}}
    \label{fig:iris_strong_first_layer_opt}
\end{figure}

\begin{table}[t]
\centering
\caption{Tolerance sweep results.}
\label{tab:sel_tol_sweep}
\begin{tabular}{|llll|}
\hline
\multicolumn{4}{|c|}{\textbf{Strongly Entangling Layer-based PQC}}                                                                                                                            \\ \hline
\multicolumn{1}{|l|}{\textbf{\begin{tabular}[c]{@{}l@{}}Tolerance\\ value\end{tabular}}} & \multicolumn{1}{l|}{\textbf{Depth}} & \multicolumn{1}{l|}{\textbf{Gate count}} & \textbf{Accuracy} \\ \hline
\multicolumn{4}{|c|}{Iris (46 depth, 240 gate count, 96\% acc)}                                                                                                                      \\ \hline
\multicolumn{1}{|l|}{\textit{0.1}}                                                       & \multicolumn{1}{l|}{38}             & \multicolumn{1}{l|}{143}                 & 93.33             \\ \hline
\multicolumn{1}{|l|}{\textit{0.01}}                                                      & \multicolumn{1}{l|}{40}             & \multicolumn{1}{l|}{161}                 & 93.33             \\ \hline
\multicolumn{1}{|l|}{\textit{0.001}}                                                     & \multicolumn{1}{l|}{41}             & \multicolumn{1}{l|}{181}                 & 93.33             \\ \hline
\multicolumn{4}{|c|}{Digits (49 depth, 300 gate count, 94.4\% acc)}                                                                                                                  \\ \hline
\multicolumn{1}{|l|}{\textit{0.1}}                                                       & \multicolumn{1}{l|}{41}             & \multicolumn{1}{l|}{159}                 & 93.05             \\ \hline
\multicolumn{1}{|l|}{\textit{0.01}}                                                      & \multicolumn{1}{l|}{41}             & \multicolumn{1}{l|}{202}                 & 91.6              \\ \hline
\multicolumn{1}{|l|}{\textit{0.001}}                                                     & \multicolumn{1}{l|}{41}             & \multicolumn{1}{l|}{222}                 & 94.4              \\ \hline
\end{tabular}
\end{table}

\textbf{Strongly entangling layer-based PQC:} To further show robustness of the proposed greedy algorithm, we optimize PQCs containing a different kind of layer called as strongly entangling layer. The main difference between basic entangler and strongly entangling layers is the parametric rotation gate. While basic entangler layer has a single rotation gate on each qubit (either RX, RY, or RZ), strongly entangling layer has a special rotation gate  R($\phi$,$\theta$,$\omega$) that is broken down in QASM format as
\[R(\phi,\theta,\omega)=RZ(\omega)\cdot RY(\theta)\cdot RZ(\phi)\]
This effectively provides the PQC with 3 times the number of tunable parameters a basic entangler layer which in turn, leads to better performance. We show the first strongly entangling layer of a pre-trained PQC trained on iris dataset in Fig. \ref{fig:iris_strong_first_layer_opt}(a).

For every strongly entangling layer of the pre-trained PQCs, we perform greedy optimization on each of the three gates present on every qubit. Fig. \ref{fig:iris_strong_first_layer_opt}(a) shows the strongly entangling layer of a PQC prior to optimization and Fig. \ref{fig:iris_strong_first_layer_opt}(b) shows the same layer after performing optimization. Similar to the case with basic entangler layers, some gates are optimized while others remain as before. In this case, most of the gates have been optimized because of a high tolerance value of 0.1. A sweep of tolerance value similar to the case of basic entangler layers was performed for strongly entangling layer-based PQCs. We select 0.1, 0.01 and 0.001 as the tolerance values and for each tolerance value, we obtain the circuit depth, gate count and accuracy of the optimized circuit. These results are tabulated in Table \ref{tab:sel_tol_sweep} and we observe minimal degradation even at high tolerance value like 0.1. This observation implies two things: (i) the strongly entangling layers are more resilient to performance degradation caused by the greedy algorithm compared to basic entangler layers, and (ii) the optimal tolerance value for strongly entangling layer-based PQCs is even higher than 0.1. Once again, by trial and error we find out that the optimal tolerance value for the two PQCs are 0.25 (iris dataset) and 0.27 (digits dataset). We tabulate the subsequent parameter values for these tolerances in Table \ref{tab:sel_opt_tol}. For iris dataset we observe around 3$\%$ reduction in accuracy for reduced gate and circuit depth after re-training, and slightly higher than 4$\%$ accuracy reduction for digits dataset. An interesting observation to note for the case of digits dataset is that re-training it at the optimal tolerance value does not improve its training accuracy. The rationale could be convergence to a local minima for optimized PQC with lower parameters.

\textbf{Comparison with existing works:} We compare the proposed greedy algorithm with the existing works as tabulated in Table \ref{tab:exist_comp}. Specifically, we compare with QuantumNAS \cite{wang2022quantumnas}, three variants of CompVQC \cite{hu2022quantum}, and knowledge distillation \cite{alam2023knowledge}. We note that nearly all the existing works outperform the proposed greedy algorithm. This might seem disadvantageous, however there are flaws in the existing works that are not present in the greedy algorithm. First, Zero-Only Pruning method in QuantumNAS and CompVQC both suffer from exponential search-space problems. As a result the search space has to be trimmed down to reduce time complexity. 
The proposed greedy algorithm has an already limited search space since the single qubit non-parametric gates in the circuit are limited and not variable. Second, all the three methods face scalability issues with respect to higher number of qubits. Furthermore, 
the proposed greedy algorithm can be easily incoroporated on top of CompVQC and knowledge distillation works to obtain extra optimization of PQCs since they do not optimize parametric gates. In the worst case, the time complexity of greedy algorithm will scale as O($n\cdot l$) for a PQC with $n$ qubits and $l$ layers in general.

\begin{table}[t]
\centering
\caption{Parameter values for optimal tolerance value}
\label{tab:sel_opt_tol}
\begin{tabular}{|llll|}
\hline
\multicolumn{4}{|c|}{\textbf{Strongly Entangling Layer-based PQC}}                                                                                                                                                                                                                                                                                                \\ \hline
\multicolumn{1}{|l|}{\textbf{}}                                                        & \multicolumn{1}{l|}{\textbf{\begin{tabular}[c]{@{}l@{}}Original \\ circuit\end{tabular}}} & \multicolumn{1}{l|}{\textbf{\begin{tabular}[c]{@{}l@{}}Optimized\\ circuit\end{tabular}}} & \textbf{\begin{tabular}[c]{@{}l@{}}Optimized \\ circuit re-trained\end{tabular}} \\ \hline
\multicolumn{4}{|c|}{Iris dataset (tolerance = 0.25)}                                                                                                                                                                                                                                                                                                            \\ \hline
\multicolumn{1}{|l|}{\textit{\begin{tabular}[c]{@{}l@{}}Accuracy\\ (\%)\end{tabular}}} & \multicolumn{1}{l|}{96}                                                                   & \multicolumn{1}{l|}{60}                                                                   & \textbf{93.33 (2.67\% red)}                                                      \\ \hline
\multicolumn{1}{|l|}{\textit{Depth}}                                                   & \multicolumn{1}{l|}{46}                                                                   & \multicolumn{1}{l|}{35}                                                                   & \textit{35 (23.9\% red.)}                                                        \\ \hline
\multicolumn{1}{|l|}{\textit{Gate count}}                                              & \multicolumn{1}{l|}{240}                                                                  & \multicolumn{1}{l|}{115}                                                                  & \textit{115 (52\% red.)}                                                         \\ \hline
\multicolumn{4}{|c|}{Digits dataset(tolerance = 0.27)}                                                                                                                                                                                                                                                                                                          \\ \hline
\multicolumn{1}{|l|}{\textit{\begin{tabular}[c]{@{}l@{}}Accuracy\\ (\%)\end{tabular}}} & \multicolumn{1}{l|}{94.4}                                                                 & \multicolumn{1}{l|}{90.27\%}                                                              & \textbf{90.27 (4.13\% red.)}                                                            \\ \hline
\multicolumn{1}{|l|}{\textit{Depth}}                                                   & \multicolumn{1}{l|}{49}                                                                   & \multicolumn{1}{l|}{32}                                                                   & \textit{32 (34.7\% red.)}                                                        \\ \hline
\multicolumn{1}{|l|}{\textit{Gate count}}                                              & \multicolumn{1}{l|}{300}                                                                  & \multicolumn{1}{l|}{128}                                                                  & \textit{128 (57.3\% red.)}                                                       \\ \hline
\end{tabular}
\end{table}

\begin{table}[t]
\centering
\caption{Comparison of proposed greedy algorithm with existing works in literature}
\label{tab:exist_comp}
\begin{tabular}{|l|l|l|}
\hline
                                                                    \textbf{\begin{tabular}[c]{@{}l@{}} Proposed work\end{tabular}}    & \textbf{\begin{tabular}[c]{@{}l@{}}\% depth \\ reduction\end{tabular}} & \textbf{\begin{tabular}[c]{@{}l@{}}\% accuracy\\ degradation\end{tabular}} \\ \hline
\begin{tabular}[c]{@{}l@{}}Zero-Only Pruning\\ (QuantumNAS \cite{wang2022quantumnas})\end{tabular} & 42.1                                                                   & 2.16                                                                       \\ \hline
CompVQC-Pruning \cite{hu2022quantum}                                                         & 38.8                                                                   & 0.91                                                                       \\ \hline
CompVQC-Quant \cite{hu2022quantum}                                                            & 10.7                                                                   & 1.75                                                                       \\ \hline
CompVQC \cite{hu2022quantum}                                                                 & 61.1                                                                   & 0.91                                                                       \\ \hline
Knowledge Distillation \cite{alam2023knowledge}                                                  & 71.4                                                                   & 5-10\%                                                                     \\ \hline
\begin{tabular}[c]{@{}l@{}}Proposed greedy\\ algorithm\end{tabular}      & 34.7                                                                   & 4.13                                                                       \\ \hline
\end{tabular}
\end{table}

\section{Discussion}
\begin{figure}[t]
    \centering
    \includegraphics[width=\linewidth]{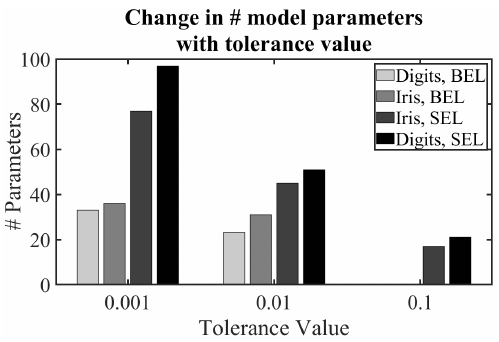}
    \caption{Change in number of parameters in optimized PQC with tolerance value. BEL: basic entangler layer, SEL: strongly entangling layer}
    \label{fig:param_vs_tol}
\end{figure}

\textbf{Tolerance value vs $\#$of remaining/non-decomposed parameters:} As the tolerance value increases, the number of remaining PQC parameters (i.e., parametric rotation angles) reduces. This is because at higher tolerance values, more number of parametric gates will be converted to their approximate non-parametric representations. We summarize the parameter trend with tolerance value in Fig. \ref{fig:param_vs_tol}. From the bar graph, we note (i) for basic entangler layer-based PQCs, digits dataset has lower number of parameters compared to iris dataset, and (ii) for strongly entangling layer-based PQC, iris dataset has lower number of parameters compared to digits dataset. For tolerance value of 0.1, both digits and iris dataset for basic entangler layer-based PQC have zero parameters left since all the gates are approximated to their non-parametric representations.

\textbf{Greedy optimization of back-to-back parametric gates:} The proposed greedy algorithm focuses on optimizing individual parametric gates. However, when back-to-back parametric gates are individually optimized, there can be a chance that the combination of final approximated non-parametric gates may cancel out to give identity operation. For example, in the final qubit of Fig. \ref{fig:iris_strong_first_layer_opt}(b), we have $t$, $id$ and $t^{\dag}$ operations, which are approximation of individual rotation gates of R(0.57,-0.04,-0.57). However, the approximated approximation can be nullified to identity operation as $t.id.t^{\dag}=t.t^{\dag}=id$. In other words, the greedy algorithm lacks awareness on combination of parametric gates. A simple solution to this would be to generate non-parametric representation of the combined back-to-back set of parametric gates instead of working on individual parametric gates.

\textbf{Considerations for coupling constraints:} As mentioned earlier, we only take decomposition of gates into account and not the coupling constraints. As a result, only parametric rotation gates are optimized and not CNOT gates. With the introduction of coupling constraints, the SWAP insertion procedure must be taken into account, leading to an increased circuit depth and gate count compared to the optimal values shown in Table \ref{tab:optimal_values}.

\section{Conclusion}
In this work, we proposed a novel greedy algorithm to optimize parametric rotation gates by generating their approximate non-parametric representation that minimizes a distance metric based on unitary transformation matrices of original and approximate gate representations. We perform this optimization on RX gates present in basic entangler layers of pre-trained PQCs and further re-train the overall optimized PQCs to recover the performance degradation due to optimization.

\section*{Acknowledgements}
We acknowledge the usage of IBM Quantum along with Pennylane for performing all the experiments. All the relevent code has been added to a GitHub Repository\footnote{GitHub repository link: \href{https://github.com/KoustubhPhalak/Greedy-PQC-Optimization}{Greedy-PQC-Optimization}}. This work is supported in parts by NSF (CNS-1722557, CNS-2129675,CCF-2210963,CCF-1718474,OIA-2040667, DGE-1723687, DGE-1821766, and DGE-2113839) and Intel’s gift.

\bibliographystyle{IEEEtran}
\bibliography{references}

\begin{thebibliography}{10}
\providecommand{\url}[1]{#1}
\csname url@samestyle\endcsname
\providecommand{\newblock}{\relax}
\providecommand{\bibinfo}[2]{#2}
\providecommand{\BIBentrySTDinterwordspacing}{\spaceskip=0pt\relax}
\providecommand{\BIBentryALTinterwordstretchfactor}{4}
\providecommand{\BIBentryALTinterwordspacing}{\spaceskip=\fontdimen2\font plus
\BIBentryALTinterwordstretchfactor\fontdimen3\font minus \fontdimen4\font\relax}
\providecommand{\BIBforeignlanguage}[2]{{%
\expandafter\ifx\csname l@#1\endcsname\relax
\typeout{** WARNING: IEEEtran.bst: No hyphenation pattern has been}%
\typeout{** loaded for the language `#1'. Using the pattern for}%
\typeout{** the default language instead.}%
\else
\language=\csname l@#1\endcsname
\fi
#2}}
\providecommand{\BIBdecl}{\relax}
\BIBdecl

\bibitem{georgescu2014quantum}
I.~M. Georgescu, S.~Ashhab, and F.~Nori, ``Quantum simulation,'' \emph{Reviews of Modern Physics}, vol.~86, no.~1, p. 153, 2014.

\bibitem{buluta2009quantum}
I.~Buluta and F.~Nori, ``Quantum simulators,'' \emph{Science}, vol. 326, no. 5949, pp. 108--111, 2009.

\bibitem{kim2023evidence}
Y.~Kim, A.~Eddins, S.~Anand, K.~X. Wei, E.~Van Den~Berg, S.~Rosenblatt, H.~Nayfeh, Y.~Wu, M.~Zaletel, K.~Temme \emph{et~al.}, ``Evidence for the utility of quantum computing before fault tolerance,'' \emph{Nature}, vol. 618, no. 7965, pp. 500--505, 2023.

\bibitem{schuld2015introduction}
M.~Schuld, I.~Sinayskiy, and F.~Petruccione, ``An introduction to quantum machine learning,'' \emph{Contemporary Physics}, vol.~56, no.~2, pp. 172--185, 2015.

\bibitem{cao2019quantum}
Y.~Cao, J.~Romero, J.~P. Olson, M.~Degroote, P.~D. Johnson, M.~Kieferov{\'a}, I.~D. Kivlichan, T.~Menke, B.~Peropadre, N.~P. Sawaya \emph{et~al.}, ``Quantum chemistry in the age of quantum computing,'' \emph{Chemical reviews}, vol. 119, no.~19, pp. 10\,856--10\,915, 2019.

\bibitem{herman2022survey}
D.~Herman, C.~Googin, X.~Liu, A.~Galda, I.~Safro, Y.~Sun, M.~Pistoia, and Y.~Alexeev, ``A survey of quantum computing for finance,'' \emph{arXiv preprint arXiv:2201.02773}, 2022.

\bibitem{ur2023quantum}
R.~Ur~Rasool, H.~F. Ahmad, W.~Rafique, A.~Qayyum, J.~Qadir, and Z.~Anwar, ``Quantum computing for healthcare: A review,'' \emph{Future Internet}, vol.~15, no.~3, p.~94, 2023.

\bibitem{saki2021survey}
A.~A. Saki, M.~Alam, K.~Phalak, A.~Suresh, R.~O. Topaloglu, and S.~Ghosh, ``A survey and tutorial on security and resilience of quantum computing,'' in \emph{2021 IEEE European Test Symposium (ETS)}.\hskip 1em plus 0.5em minus 0.4em\relax IEEE, 2021, pp. 1--10.

\bibitem{zhang2020recent}
Y.~Zhang and Q.~Ni, ``Recent advances in quantum machine learning,'' \emph{Quantum Engineering}, vol.~2, no.~1, p. e34, 2020.

\bibitem{sun2019survey}
S.~Sun, Z.~Cao, H.~Zhu, and J.~Zhao, ``A survey of optimization methods from a machine learning perspective,'' \emph{IEEE transactions on cybernetics}, vol.~50, no.~8, pp. 3668--3681, 2019.

\bibitem{schuld2020circuit}
M.~Schuld, A.~Bocharov, K.~M. Svore, and N.~Wiebe, ``Circuit-centric quantum classifiers,'' \emph{Physical Review A}, vol. 101, no.~3, p. 032308, 2020.

\bibitem{cerezo2021cost}
M.~Cerezo, A.~Sone, T.~Volkoff, L.~Cincio, and P.~J. Coles, ``Cost function dependent barren plateaus in shallow parametrized quantum circuits,'' \emph{Nature communications}, vol.~12, no.~1, p. 1791, 2021.

\bibitem{killoran2019continuous}
N.~Killoran, T.~R. Bromley, J.~M. Arrazola, M.~Schuld, N.~Quesada, and S.~Lloyd, ``Continuous-variable quantum neural networks,'' \emph{Physical Review Research}, vol.~1, no.~3, p. 033063, 2019.

\bibitem{abbas2021power}
A.~Abbas, D.~Sutter, C.~Zoufal, A.~Lucchi, A.~Figalli, and S.~Woerner, ``The power of quantum neural networks,'' \emph{Nature Computational Science}, vol.~1, no.~6, pp. 403--409, 2021.

\bibitem{du2020expressive}
Y.~Du, M.-H. Hsieh, T.~Liu, and D.~Tao, ``Expressive power of parametrized quantum circuits,'' \emph{Physical Review Research}, vol.~2, no.~3, p. 033125, 2020.

\bibitem{dawson2005solovay}
C.~M. Dawson and M.~A. Nielsen, ``The solovay-kitaev algorithm,'' \emph{arXiv preprint quant-ph/0505030}, 2005.

\bibitem{von2018mathematical}
J.~Von~Neumann, \emph{Mathematical foundations of quantum mechanics: New edition}.\hskip 1em plus 0.5em minus 0.4em\relax Princeton university press, 2018, vol.~53.

\bibitem{alam2023knowledge}
M.~Alam, S.~Kundu, and S.~Ghosh, ``Knowledge distillation in quantum neural network using approximate synthesis,'' in \emph{Proceedings of the 28th Asia and South Pacific Design Automation Conference}, 2023, pp. 639--644.

\bibitem{hinton2015distilling}
G.~Hinton, O.~Vinyals, and J.~Dean, ``Distilling the knowledge in a neural network,'' \emph{arXiv preprint arXiv:1503.02531}, 2015.

\bibitem{hu2022quantum}
Z.~Hu, P.~Dong, Z.~Wang, Y.~Lin, Y.~Wang, and W.~Jiang, ``Quantum neural network compression,'' in \emph{Proceedings of the 41st IEEE/ACM International Conference on Computer-Aided Design}, 2022, pp. 1--9.

\bibitem{aleksandrowicz2019qiskit}
G.~Aleksandrowicz, T.~Alexander, P.~Barkoutsos, L.~Bello, Y.~Ben-Haim, D.~Bucher, F.~J. Cabrera-Hern{\'a}ndez, J.~Carballo-Franquis, A.~Chen, C.-F. Chen \emph{et~al.}, ``Qiskit: An open-source framework for quantum computing,'' \emph{Accessed on: Mar}, vol.~16, 2019.

\bibitem{mark_skilbeck_2020_3967926}
\BIBentryALTinterwordspacing
M.~Skilbeck, E.~Peterson, appleby, E.~Davis, P.~Karalekas, J.~M. Bello-Rivas, D.~Kochmanski, Z.~Beane, R.~Smith, A.~Shi, C.~Scott, A.~Paszke, E.~Hulburd, M.~Young, A.~S. Jackson, BHAVISHYA, M.~S. Alam, W.~Velázquez-Rodríguez, c.~b. osborn, fengdlm, and jmackeyrigetti, ``rigetti/quilc: v1.21.0,'' Jul. 2020. [Online]. Available: \url{https://doi.org/10.5281/zenodo.3967926}
\BIBentrySTDinterwordspacing

\bibitem{sivarajah2020t}
S.~Sivarajah, S.~Dilkes, A.~Cowtan, W.~Simmons, A.~Edgington, and R.~Duncan, ``t| ket>: a retargetable compiler for nisq devices,'' \emph{Quantum Science and Technology}, vol.~6, no.~1, p. 014003, 2020.

\bibitem{li2023quarl}
Z.~Li, J.~Peng, Y.~Mei, S.~Lin, Y.~Wu, O.~Padon, and Z.~Jia, ``Quarl: A learning-based quantum circuit optimizer,'' \emph{arXiv preprint arXiv:2307.10120}, 2023.

\bibitem{fosel2021quantum}
T.~F{\"o}sel, M.~Y. Niu, F.~Marquardt, and L.~Li, ``Quantum circuit optimization with deep reinforcement learning,'' \emph{arXiv preprint arXiv:2103.07585}, 2021.

\bibitem{moflic2023graph}
I.~Moflic, V.~Garg, and A.~Paler, ``Graph neural network autoencoders for efficient quantum circuit optimisation,'' \emph{arXiv preprint arXiv:2303.03280}, 2023.

\bibitem{zhou2020monte}
X.~Zhou, Y.~Feng, and S.~Li, ``A monte carlo tree search framework for quantum circuit transformation,'' in \emph{Proceedings of the 39th International Conference on Computer-Aided Design}, 2020, pp. 1--7.

\bibitem{duncan2020graph}
R.~Duncan, A.~Kissinger, S.~Perdrix, and J.~Van De~Wetering, ``Graph-theoretic simplification of quantum circuits with the zx-calculus,'' \emph{Quantum}, vol.~4, p. 279, 2020.

\bibitem{wang2022quantumnas}
H.~Wang, Y.~Ding, J.~Gu, Y.~Lin, D.~Z. Pan, F.~T. Chong, and S.~Han, ``Quantumnas: Noise-adaptive search for robust quantum circuits,'' in \emph{2022 IEEE International Symposium on High-Performance Computer Architecture (HPCA)}.\hskip 1em plus 0.5em minus 0.4em\relax IEEE, 2022, pp. 692--708.

\bibitem{misc_iris_53}
R.~A. Fisher, ``{Iris},'' UCI Machine Learning Repository, 1988, {DOI}: https://doi.org/10.24432/C56C76.

\bibitem{misc_optical_recognition_of_handwritten_digits_80}
E.~Alpaydin and C.~Kaynak, ``{Optical Recognition of Handwritten Digits},'' UCI Machine Learning Repository, 1998, {DOI}: https://doi.org/10.24432/C50P49.

\end{thebibliography}

\end{document}